# Analysis and Calibration of Electron-Dispersive Spectroscope and Scanning Electron Microscope Parameters to Improve their Results

Hamidreza Moradi[1*], Fatemeh Mehradnia[2]

[1]Department of Mechanical Engineering and Engineering Science, The University of North Carolina at Charlotte, Charlotte, North Carolina, USA

[1]Center for Translational Medicine, Department of Biomedical and Pharmaceutical Sciences, University of Montana, USA

[*] Corresponding Author

**Abstract**

The Scanning electron microscope (SEM) and Electron-Dispersive Spectroscope (EDS) are two highly effective instruments in the field of nanoscience and nanotechnology. The quality of these instruments is determined by various factors, with high resolution being a crucial one. To determine the practically achievable resolution of the SEM, reference materials such as gold particles on carbon substrate are commonly utilized. On the other hand, high demanding, and usual materials such as brass, aluminum oxide or MEMs are of high importance to be considered in the way of improving material characterization. In this study five different samples were analyzed in four steps to see the effect of astigmatism and aperture misalignment, the mirror effect on the charged uncoated sample in backscattered electron (BSE) and secondary electron (SE) images, compositional and topographical analysis using both qualitative and quantitative methods and finally apply all the steps on a micro-electromechanical system (MEMS) device.  This study illustrates the possibility of controlling the mirror effect by adjusting the input parameters even in the cases that the user needs to have low accelerating voltage for the experiment. Besides, the discussion addresses possible reasons for the absence of some peaks in EDS analysis for some specific elements.

## 1. Introduction

In the field of nanoscience and engineering, it is essential to characterize materials. The SEM technique can be used to image any substrate, whether it is conducting, insulating, or biological material, provided that the sample preparation process is appropriate [1]. Accurate measurements are required for samples with sizes smaller than 5 nm, such as carbon ultra-thin films, gold nanoparticles, impurities, or quantum dots. Precise size estimation is crucial because new properties may emerge depending on the shape and size of nanomaterials. Some SEM manufacturers claim that their instruments have a resolution of less than 1 nm, which means they can distinguish two closely spaced particles as separate entities in the SEM image. However, when purchasing an SEM instrument, it is important to determine whether it can resolve ultra-small objects. Unfortunately, commercially available traceable reference materials or specimens for determining SEM resolution do not exist yet.

One can create samples by growing localized gold nanoparticles (Au NPs) on user-friendly substrates like $SiO_2$/Si. To determine the resolution limit of the SEM instrument, one can measure the minimum distance between two nanoparticles using the gap method. Other methods like derivative, fast Fourier transform, and contrast to gradients are available but not widely used [2-4]. Gold nanoislands on such substrates are considered the industry standard for testing SEM instrument resolution but require international acceptance due to the lack of traceable resolution test specimen standards and accepted guidelines [5]. In addition to being used for SEM resolution specimens, Au NPs are also important for various applications that require signal amplification such as biological tags, chemical and optical sensors, and enhanced surface plasmon resonance generation [6-8].

The element carbon is recognized for emitting low levels of secondary electrons when compared to copper [9-13]. Researchers have conducted an experiment to examine the impact of a thin layer of carbon deposited on a copper substrate [13]. and the results show that even a very thin layer of carbon can significantly reduce electron emission from the copper substrate. For this study, the first sample will consist of Au particles on $SiO_2$/Si substrate and Sample and Cu grid with ultra-thin C.

The food packaging and flat panel display industries place great importance on the microstructural and gas barrier properties of transparent metal oxide layers deposited on plastic substrates. These ceramic layers can be customized to offer various properties such as transparency, microwave compatibility, conductivity, and environmental safety. However, defects like nanoscale pores and micron-scale pinhole defects limit the gas barrier properties of these oxide films [14-18]. While silicon oxide coatings on poly Žethylene terephtha- late. ŽPET have been extensively studied as a gas barrier system [16-20], other inorganic coatings have not received similar attention except for aluminum oxide. In this study, $Al_2O_3$ powder dispersed on a plastic substrate with a thin Au layer on top will be used as the second sample for machine calibration purposes.

Brass is an alloy of copper and zinc that can have varying compositions. Although it has been used for centuries, its metallurgy was not fully understood until the 18th century [21-24]. Alpha-brass, which contains more than 65% copper, is a soft and ductile alloy that was commonly used in ancient times. The solubility of zinc in copper can be as high as 40% [25, 26]. The addition of zinc improves brass's corrosion resistance due to the formation of complex passive oxides on the metal's surface. α-brass has better corrosion resistance compared to pure copper [27, 28]. Lead may also be present in brass alloys due to the use of lead-rich ore [29].

Microelectromechanical systems (MEMS) have great potential for integrating sensors, actuators, and electronics into a small package. However, their small size presents engineering difficulties such as stiction, friction, and wear. Most MEMS are made from polysilicon due to its ease of batch processing in the semiconductor industry, but polysilicon-on-polysilicon contacts have poor friction and wear properties [30-34]. Sometimes, the experimental design is guided by various numerical analyses [35, 36]. In this work, all processes will be performed on a MEMS device at different accelerating voltages and spot sizes.

Compositional and topographical investigation of different samples are currently being caried out using scanning electron microscope (SEM) along with Electron-Dispersive Spectroscopy (EDS). Due to low cost and various applications SEM is the most popular method among other electron-based analysis methods. This study can be a great help in different fields such as, determining the crack propagation direction [37], volume of the dead cells in biological parts [38] and elemental compositional analysis of different samples after the production of novel samples [39]{Modaresahmadi, 2023 #2}{Modaresahmadi, 2022 #1} are some of the examples from limitless application of these method. However, setting proper parameters in the experiment (e.g., accelerating voltage, spot size, working distance, aperture size etc.) and understanding the result of any changes in the SEM (e.g., charging, shadowing, mirroring effect, etc.) and EDS parameters is a challenge scientist working with these methods.

In this study five different samples were analyzed in four steps each of which study different phenomena. In the first step two samples were analyzed to see the effect of astigmatism and aperture misalignment. In the second step coating procedure of a sample was caried out following by SEM analysis to study the difference of coated and uncoated materials as well as backscattered electron (BSE) and secondary electron (SE) images. In this step we see the mirror effect on the charged uncoated sample. In the third step we conduct a compositional and topographical analysis of a sample using both qualitative and quantitative methods. To evaluate the previous results, in the last step we apply all the steps on a micro-electromechanical system (MEMS) device.

## 2. Methodology, Experiment and Procedure

In this work JEOL JSM-6480 SEM was used to conduct the Scanning Electron Microscopy analysis and Denton Desk IV was used to coat the samples. We use different samples in three different steps for which the procedures and settings which will be discussed in detail.

### 2.1 *Au* and *Cu* samples

In the first step we analyze two samples Sample 1: Au particles on $SiO_2$/Si substrate and Sample.
2: Cu grid with ultra-thin C.

#### Sample preparation

We need to have a conductive sample to perform the experiment which will usually been done by a coating process. In case of having metallic sample there is no need to have coating on the sample.

#### Operation and alignment

Considering coordinates shown in figure 1 and based on the fact that the sample height is small and working distance ($W_D$) is 10mm, we set Z to be 10mm. Notably, as the working distance increases the resolution of the image enhances and the prob size decreases.

As he acceleration voltage ($V_{ac}$) increases the results would be better however it is possible that the sample heat up and get damaged. For metallurgical samples accelerating voltage between 15-30V is recommended, therefore, it is set to be 20kV. As spot size decrease, we have better resolution however there will be less current and lesser signal to noise ratio causing the image to be blurry, so we need to have a balance in the value. Hence, the spot size (SP) is set to be 30.

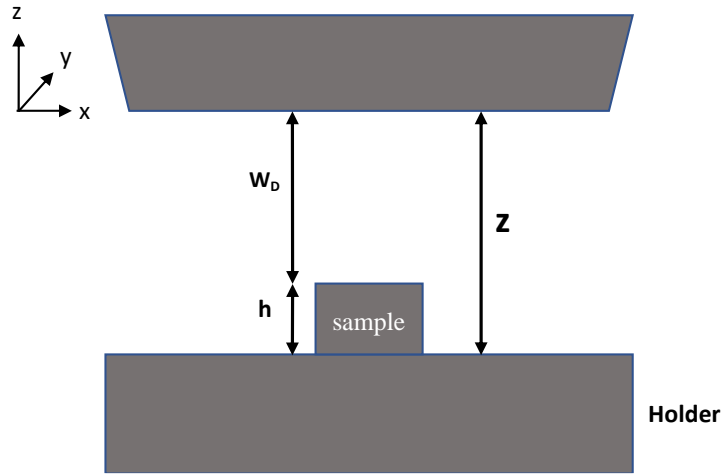

**Figure 1.** Position of the sample in the SEM

At the next step, we use coarse shift to find the sample and then use fine shift to pick a spot on the sample. Then, we increase magnification to see some features on the surface (the image would not be clear) and then using coarse focus we enhance the focus of the image. We increase the magnification again and using fine focus we enhance the image.

To adjust the objective lens aperture, we set the magnification on ×20000 and choose OL wobbler which shift the image to scan 1. Using X/Y directions on the aperture, we try to minimize image shifting in each direction.

Using over-focused and under-focused images, we check to see if we have any stretching in our objects in the image. In case of any stretching, we need to adjust the focused image to get the sharpest image.

Next, we decrease magnification to ×3000 and we move the beam over TEM grid (which is Cu grid with ultra-thin C), coating. In an attempt we use different Vac at SP=30 and in another attempt, we use different SP at Vac=5kv to see the effect of each parameter on the results.

### 2.2 $Al_2O_3$ on plastic substrate

#### Coating procedure

The goal of this experiment is to coat partially the non-conductive $Al_2O_3$ powder dispersed on a plastic substrate with thin Au layer on top of it using Denton Desk IV sputter. We use simple shadow mask to partially coat the sample to see the difference of coated and uncoated materials. Using two pumps, we drop the pressure of the chamber from 967mTorr to a few mTorr (6 mTorr) and we use Argon gas molecules to bombard on the sample since it doesn't interact with the material.

We use radio frequency (RF) sputtering to ionize the gas molecules which using the Lorentz force change the path of electron from cathode to anode leads to increase the chance of hitting the target. The rotation setpoint is set to be 0-30% to prevent shadow effect. For the calibrated tooling factor, we follow the standard for gold.

The thickness of $Al_2O_3$ is 1μm and Au coating is 11nm. We seek to see the difference between coated and uncoated surfaces, and we also can see the results of an extremely charged surface.

### SEM analysis of the sample

We set Z to 15mm and accelerating voltage 20 KV and spot size 30. Following the procedure described above, we adjust the astigmatism and aperture misalignment. The magnification used for each image is illustrated at the bottom of the images. We stay a while on the uncoated surface to make it charged and then we change the accelerating voltage to 5KV.

## 2.3 Brass sample

The goal of this experiment is to get BSE images and EDS spectrum of the Brass sample using Oxford EDS system. We set Z to 30mm since the thickness of the sample is about 16mm. There are always some impurities inside the Brass sample, and we seek to detect the elements inside the sample.

### EDS analysis

For x-ray EDS analysis, we set the aperture on number 3, and we have the spectrum range (0-20) KeV. Moreover, we increase the spot size from 30 to 60 to reduce the dead time (the percentage of signal rejection).

## 2.4 MEMS device

The goal of this experiment is to perform all the abovementioned processes on a MEMS device at different accelerating voltages, spot size, comparing SE and BSE, comparing Compositional and topographic images and EDS analysis of the elements.

# 3. Results and Discussion

## 3.1 Aperture misalignment

In figure 2 we see the aperture effect in both x and y direction for Au particles on SiO2/Si substrate. As we observe, the grains have shifting in both directions which shows misalignment of aperture.

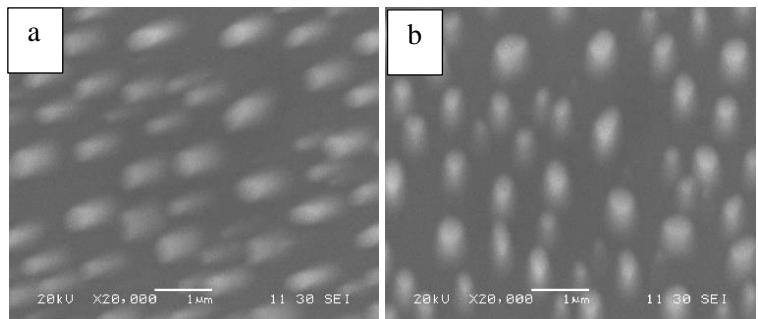

**Figure 2.** Effect of misaligned objective lens aperture in (a) x and (b) y-direction for Au particles on SiO2/Si substrate

## 3.2 Astigmatism

In figure 3 we see effect of astigmatism for all under-focused, over-focused and focused modes for Au particles on SiO2/Si substrate. In this figure we see that the grains are stretched in two perpendicular directions in over-focused and under-focused modes.

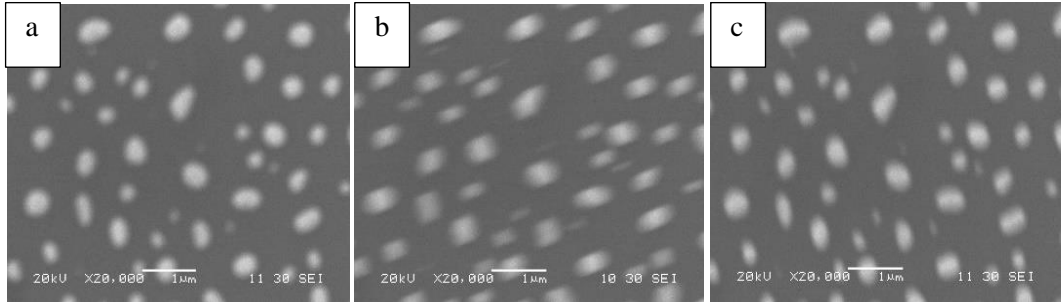

**Figure 3.** Astigmatism misaligned image in (a) focused (b) over-focused and (c) under-focused modes for Au particles on $SiO_2$/Si substrate.

In figure 4 we see that there is no stretching in the grains which shows resolving astigmatism issue for the Au particles on SiO2/Si substrate.

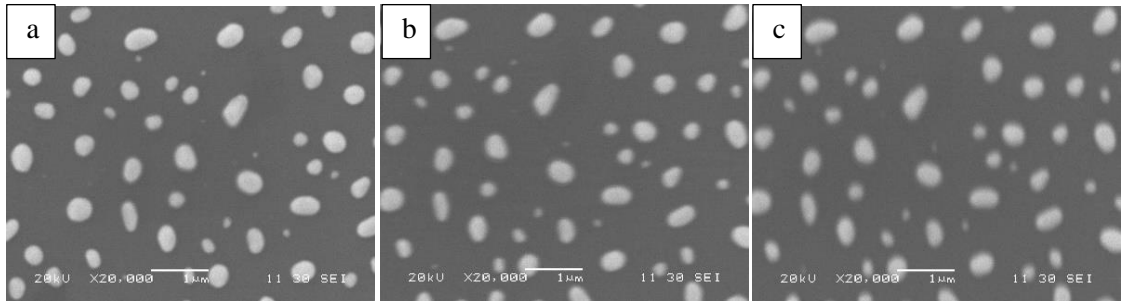

**Figure 4.** Aligned image in (a) focused (b) over-focused and (c) under-focused modes for Au particles on $SiO_2$/Si substrate.

## 3.3 Accelerating voltage effect

In figure 5 we see some small white spots on the Cu grid with ultra-thin C sample, that is related to the surface of it. However, we don't see those spots in higher voltages. This phenomenon is because of low voltage sensitivity to the surface properties.

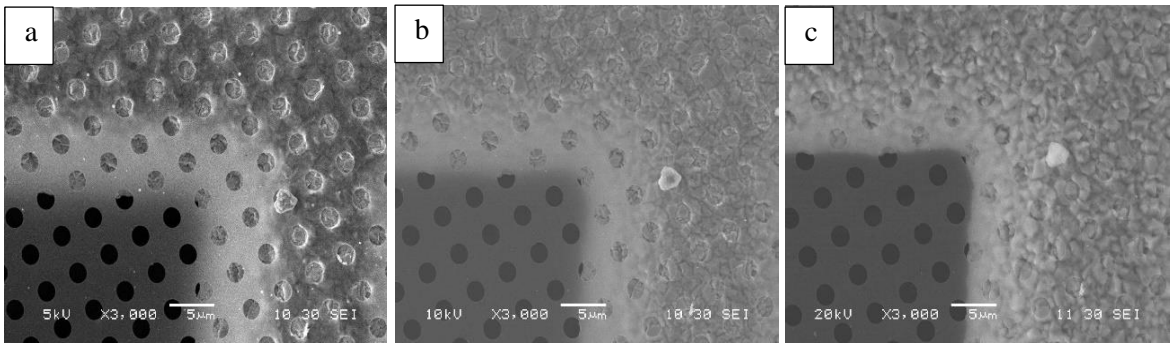

**Figure 5.** Cu grid with ultra-thin C sample under accelerating voltage of (a) 5kv (b) 10kv and (c) 20kv at the same spot size 30

### 3.4 Spot size effect

Considering figure 6, as the spot size increase, the resolution decrease however we have more signals. At large spot size we still can get high resolution by having low voltage and low magnification. In this figure we observe that at SP 60 the edge is not very sharp, and the resolution is deteriorated.

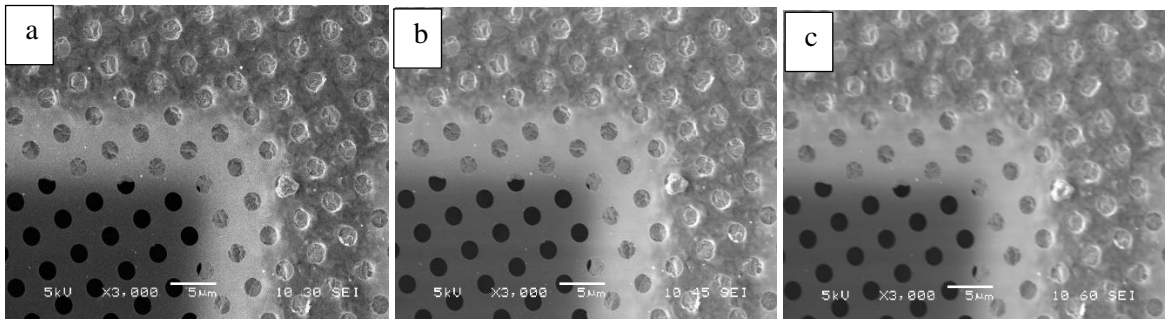

**Figure 6.** Cu grid with ultra-thin C sample at spot size of (a) 30 (b) 45 and (c) 60 at same accelerating voltage of 5kv

### 3.5 Brightness of the coated and uncoated sample

In figure 7 we see the difference between coated and uncoated areas of $Al_2O_3$ powder dispersed on a plastic substrate. Coated area results in darker image compared to uncoated area because of the charging effect. During the experiment we observed movement of the powder in uncoated area which is because of the charging effect.

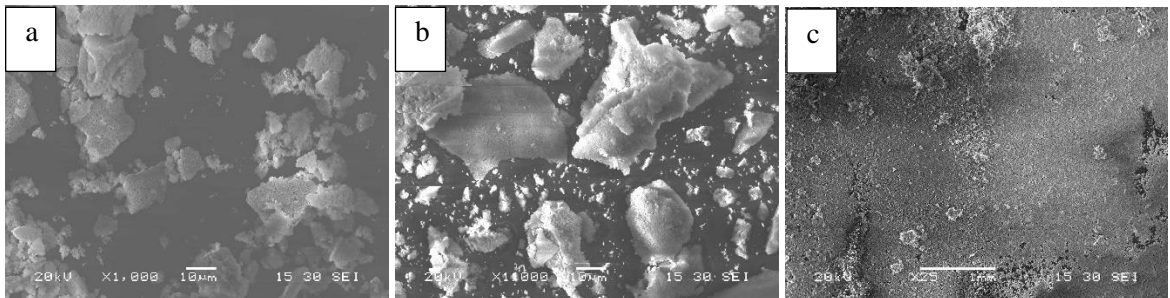

**Figure 7.** $Al_2O_3$ powder on a plastic substrate sample (a) coated (b) uncoated (c) boundary area

### 3.6 Extremely charged and uncoated sample image

In figure 8 we see the extreme charging effect of the $Al_2O_3$ powder sample using BSE images. The BSE detector includes 5 different segments, four of which make a circle shape and the other part is separated. The BSE can show surface topography named as Topo, material composition called Compo and the combination of them called Shadow mode. In compo mode since the signal is A+B both parts are illustrated as bright segments and we cannot see C part. In Topo mode, since it receives A-B the left part is illustrated as a black segment, and we don't see C again. The shadow mode has a range from 1 as the highest to 10 as the lowest coefficient of A. In shadow 1 mode we have A+B+C we see all the parts in bright and in shadow 10 mode we see that the left part is a little darker since we have a difference between A and B in their signals.

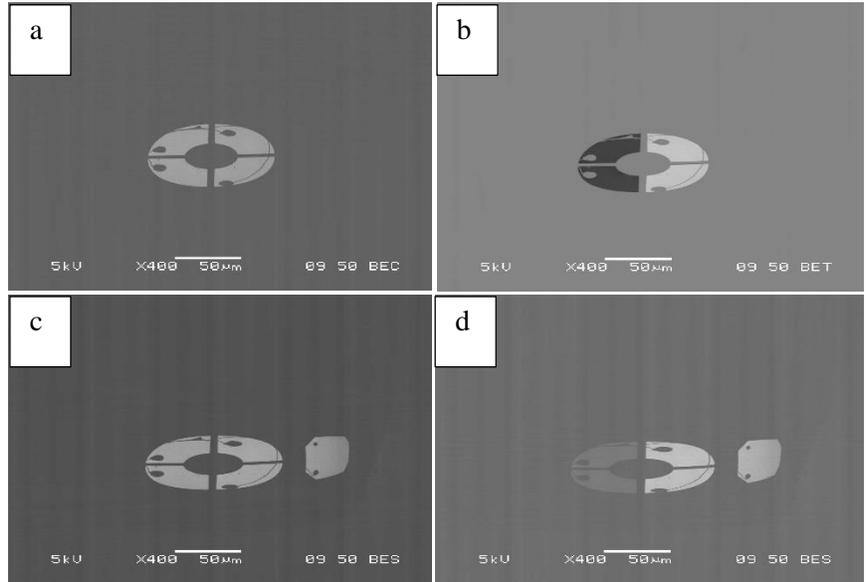

**Figure 8.** Charged BSE image of uncoated area in (a) compo mode (b) topo mode (c) shadow 1 mode and (d) shadow 10 mode.

In figure 9 we see the SE image of the uncoated charged $Al_2O_3$ powder sample. The center area of the sample acts like a charged pond which reflect the electron beam back which is called mirror effect and we see the head of the SEM machine in the image.

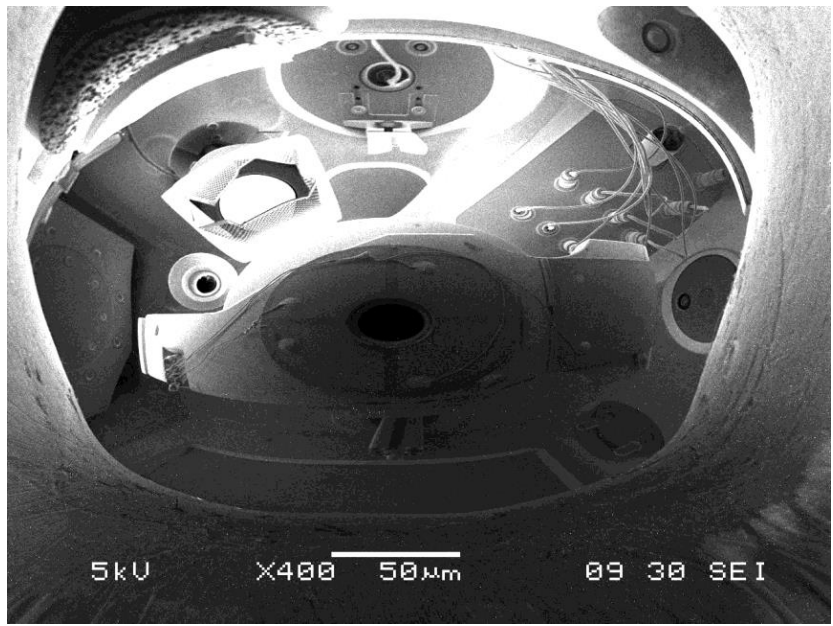

### 3.7 BSE analysis of impurities

SE is not sensitive to atomic number, but BSE is sensitive to atomic number. Figure 10 shows the surface morphology of the Brass sample. Considering figure 11 we see that in compo mode we have several bright particles which shows that they are heavier than the matrix. In topo mode we see that all the particles have the same color since it shows the surface topography of the sample and shows the slope change in the sample. In shadow mode we see the combination of compositional and topographical mode together and it illustrate a 3-D view from the sample. To determine the exact element in the sample we need to use EDS to analyze the elements. As we observe in the following figures, the bright particle in SE image is illustrated as a dark particle in BSE.

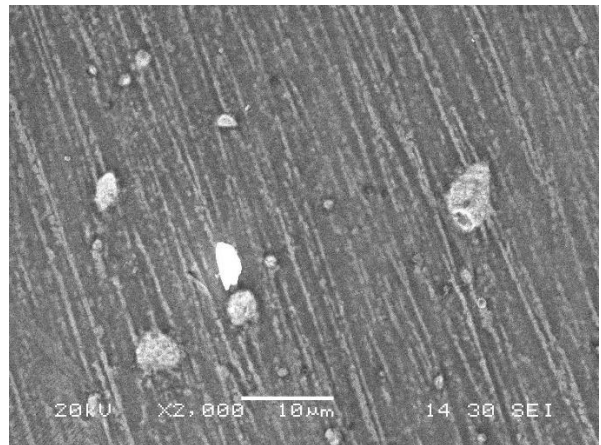

**Figure 10.** SE image of Brass sample

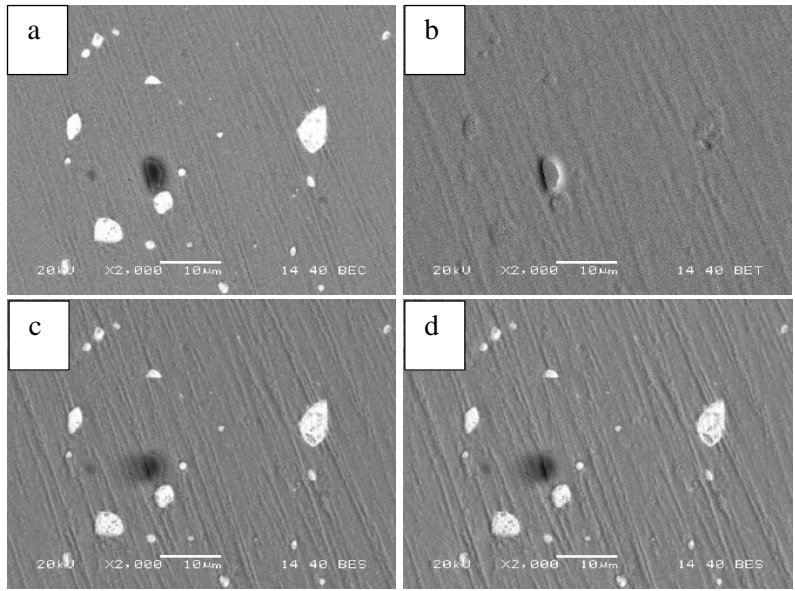

**Figure 11.** BSE image of Brass sample in (a) Compo (b) Topo (c) Shadow 1 (d) Shadow 10 mode

### 3.8 EDS analysis of impurities

In figure 12 we see the qualitative EDS analysis of Brass sample which include impurities. Considering the BSE image, we know that the bright particle at center should be a light element which is also approved in EDS qualitative analysis. We see that the bright particle is oxygen (yellow), most of the impurities are Lead (purple), the matrix is composed of Zink and Copper (red and green) and we also have some Silicon as impurities.

In figures 13-16 we see the x-ray spectrum for point identification of Brass sample which is the combination of Zink and Copper for Lead particles, matrix(the material except the impurities), overall (matrix and impurities) and dust, respectively. In each spectrum we see the elements inside the sample as a separate peak, however, some peaks are not illustrated in the spectrum because of the fact that it is deep inside the material.

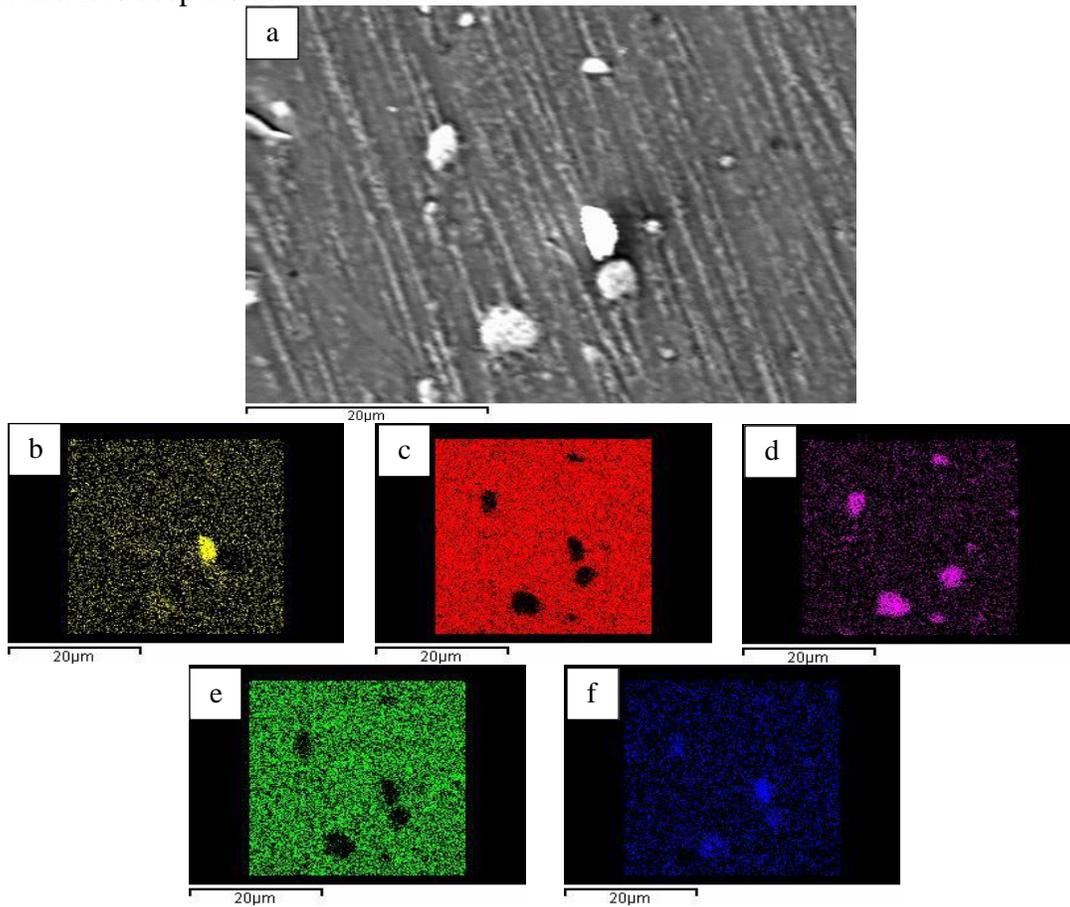

**Figure 12.** Brass sample (a) overall (b) O (c) Cu (d) Pb (e) Zn and (f) Si mapping image

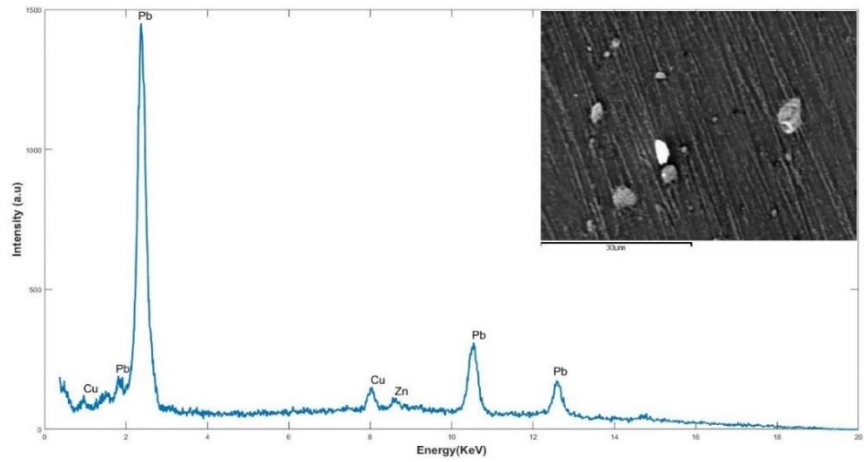

**Figure 13.** Spectrum of Lead particles

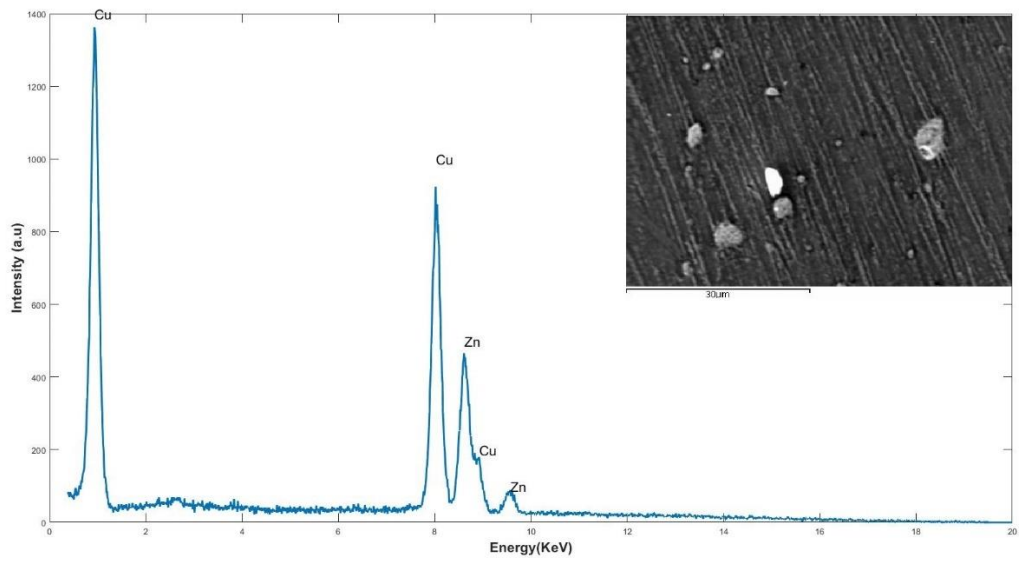

**Figure 14.** Spectrum for overall matrix

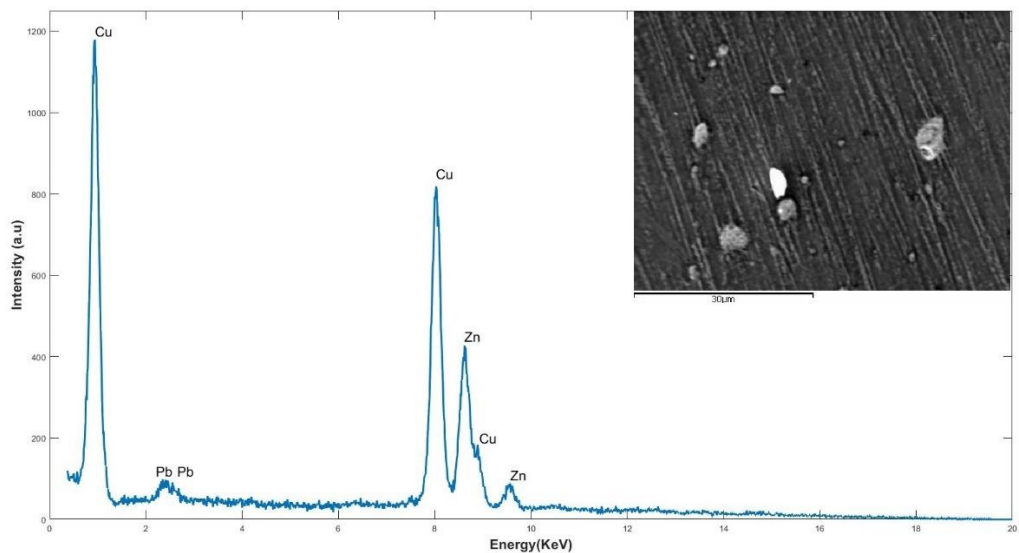

**Figure 15.** Spectrum for overall

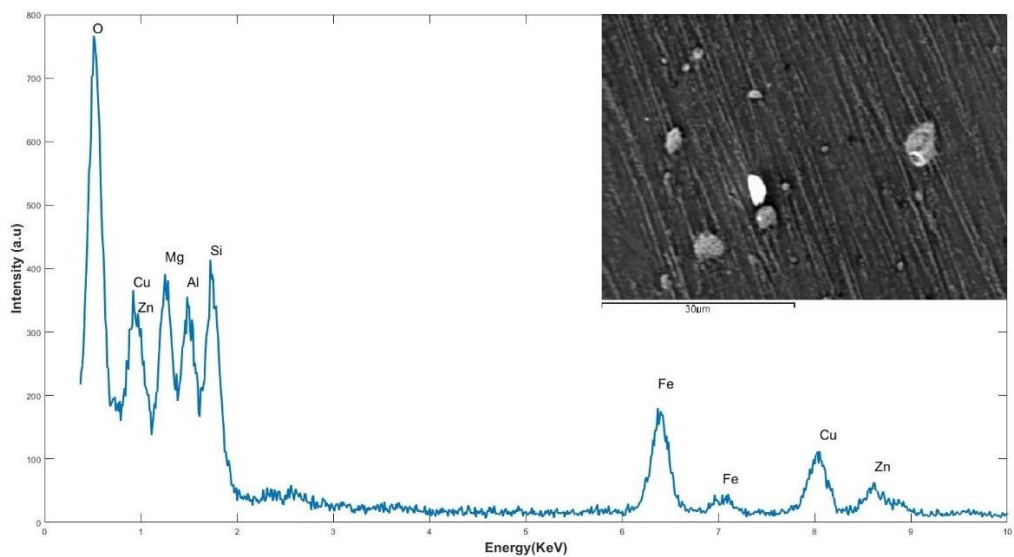

**Figure 16.** Spectrum of dust

## 3.9 Comprehensive results of MEMS device

Accelerating voltage shows the power of the electron beam penetrating through the specimen. As discussed in section 3.3, at higher accelerating voltage the sensitivity to the surface morphology decreases which can also be seen in figure 17.

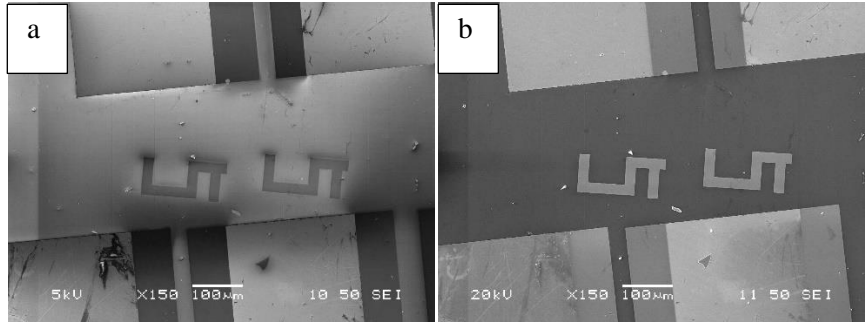

**Figure 17.** MEMS device under accelerating voltage of (a) 5kv and (b) 20kv at the same spot size 50

Figure 18 agree well with the results shown in section 3.7, showing that the compo mode features atomic number differences in the interrogated area however, topo mode illustrate the surface characteristics of the sample. We see that the shadow mode shows a better view of the sample just like a 3-D image.

Figure 19 shows the colorized mapping image of the MEMS device by which different elements of within the part can be detected. As illustrated, the S shape is Platinum and Oxygen, the bright area is Platinum, and the dark area is a combination of Silicon and Oxygen.

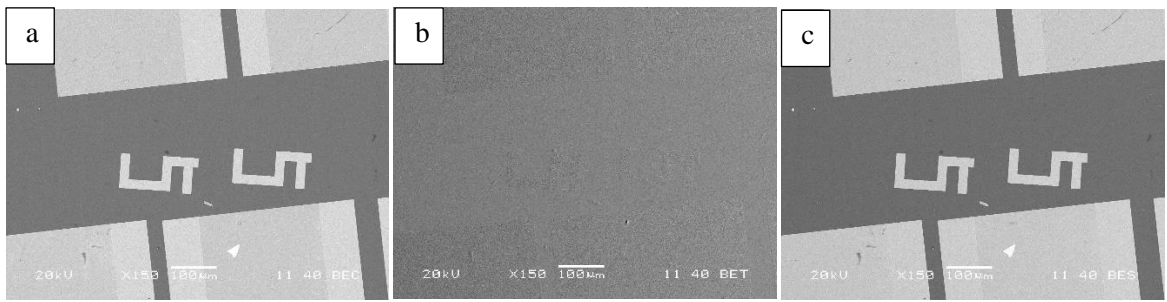

**Figure 18.** BSE image of MEMS device in (a) Compo (b) Topo and (c) Shadow at 20 KV

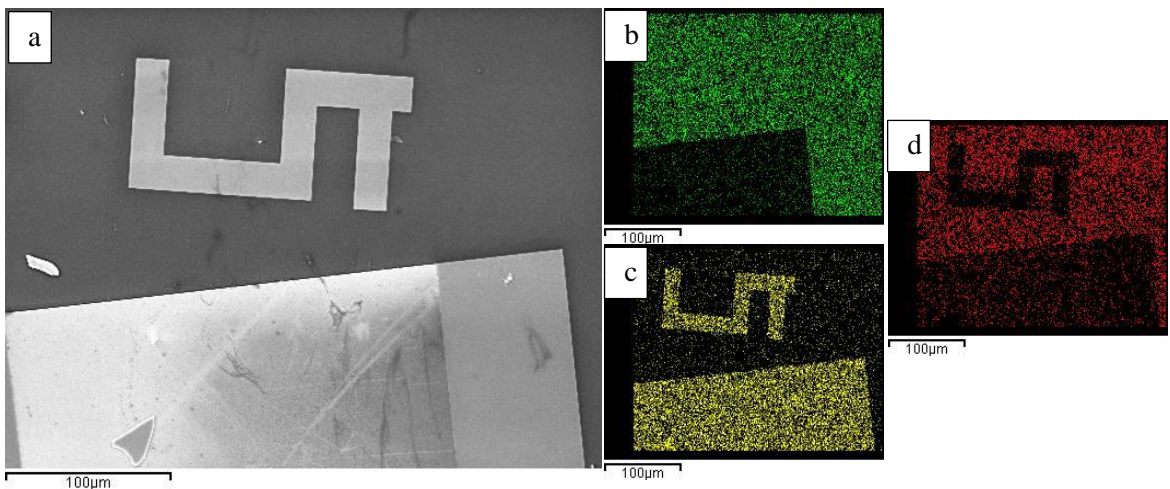

**Figure 19.** MEMS device (a) SEM (b) O (c) Pt and (d) Si mapping image

High voltage incident beam can excite the electrons from outer shells of a material which can be followed by moving the hole position from outer layer to upper layer result in x-ray generation. Figure 20-22 show the characteristic x-ray analysis of the MEMS device, showing the elements inside the material. As it can be seen, the initial part of the spectrum is removed from the chart, since it does not show correct peaks in the spectrum. We see that the corresponding x-ray images at the corner of each spectrum, has lower quality than SE and BSE images. As it can be seen from the figures below, we have Silicon and Oxygen in the blank substrate. Figure 21 shows that in the Pt coated area we have Silicon, Oxygen and Platinum present. We also see the elemental results for Pt open area in figure 22.

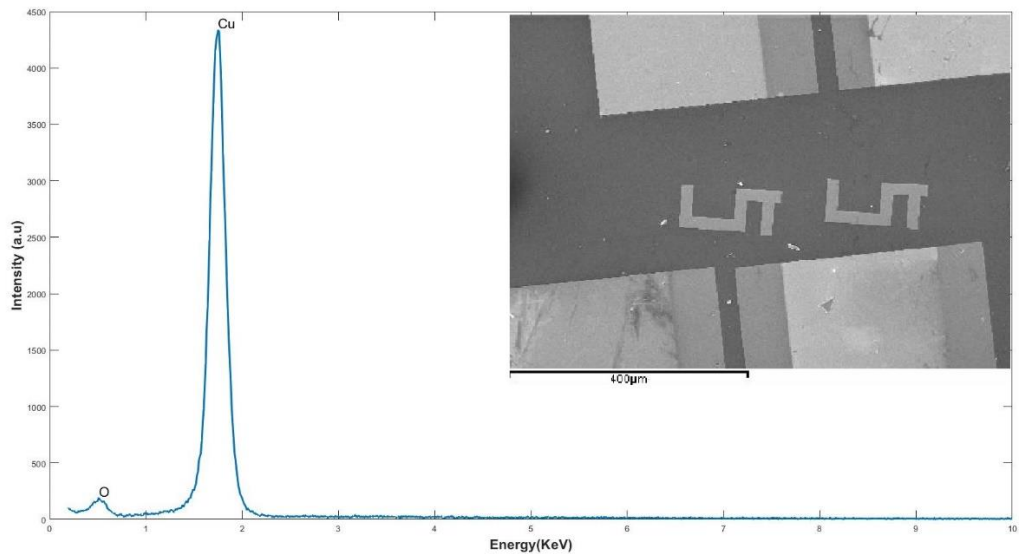

**Figure 20.** X-ray spectrum of blank substrate and the area x-ray image

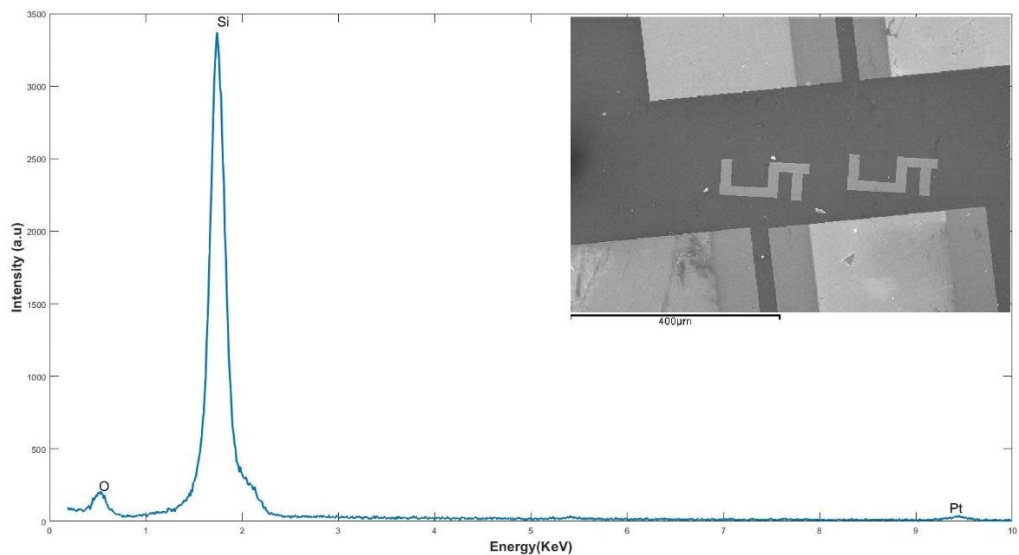

**Figure 21.** X-ray spectrum of Pt coated area and the area x-ray image

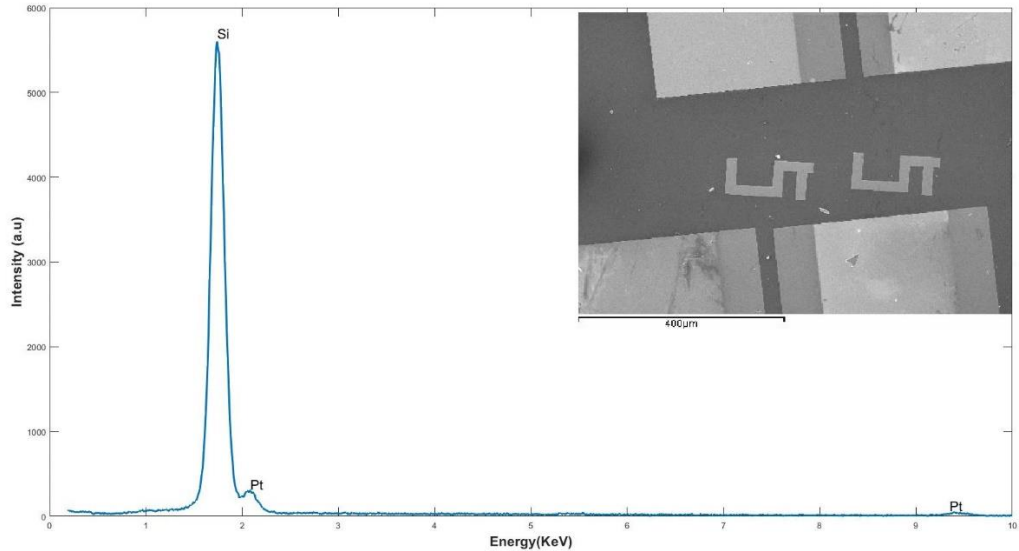

**Figure 22.** X-ray spectrum of Pt open area with the x-ray image

## 4. Conclusions

In this study which is carried out using five different samples, first the effect of astigmatism and aperture misalignment were analyzed using SEM images. Then, the role of thin coating on an unconducive sample was investigated using SE and BSE signals. The mirror effect phenomena occur for the extremely charged uncoated sample. In the next step compositional and topographical analysis of a sample using both qualitative and quantitative methods. In the last step, the evaluation of all previous phenomena was caried out using a MEMS device.

Using the outcomes of this study, researchers can adjust the SEM and EDS parameters to control undesired results (i.e., charging effect, mirror effect)